\documentstyle[12pt,aasms4]{article}
\input{psfig.tex}

\begin{document}

\title{$BVRI$ Light Curves for 29 Type Ia Supernovae}

\author{Mario Hamuy}
\affil{National Optical Astronomy Observatories$^1$, Cerro Tololo
  Inter-American Observatory, Casilla 603, La Serena, Chile; email: mhamuy@as.arizona.edu}
\affil{University of Arizona, Steward Observatory, Tucson, Arizona,
  85721}
\author{M.M. Phillips, Nicholas B. Suntzeff, R.A. Schommer}
\affil{National Optical Astronomy Observatories$^1$, Cerro Tololo
  Inter-American Observatory, Casilla 603, La Serena, Chile; emails: mphillips@noao.edu,
  nsuntzeff@noao.edu, rschommer@noao.edu}
\author{Jos\'{e} Maza$^2$, Roberto Antezana, Marina Wischnjewsky}
\affil{Departamento de Astronom\'{i}a, Universidad de Chile, Casilla 36-D,
  Santiago, Chile; email: jose@calan.das.uchile.cl}
\author{Geraldo Valladares, C\'{e}sar Muena, L. E. Gonz\'{a}lez, R. Avil\'{e}s, L.A. Wells$^3$,
R. C. Smith, Mauricio Navarrete, Ricardo Covarrubias, G. M. Williger$^4$, Alistair R. Walker,
A. C. Layden, J. H. Elias, J. A. Baldwin, Manuel Hern\'{a}ndez, Hern\'{a}n Tirado, Patricio Ugarte,
R. Elston, Nelson Saavedra}
\affil{National Optical Astronomy Observatories$^1$, Cerro Tololo
  Inter-American Observatory, Casilla 603, La Serena, Chile}
\author{Felipe Barrientos, Edgardo Costa, Paulina Lira, M. T. Ruiz, Claudio Anguita, Ximena G\'{o}mez,
P. Ortiz}
\affil{Departamento de Astronom\'{i}a, Universidad de Chile, Casilla 36-D,
  Santiago, Chile}
\author{M. Della Valle, J. Danziger, J. Storm}
\affil{La Silla Observatory, European Southern Observatory, Casilla 19001, Santiago, Chile}
\author{Y-C. Kim, C. Bailyn, E. P. Rubenstein, D. Tucker$^5$, S. Cersosimo, R. A. M\'{e}ndez$^6$,
L. Siciliano, W. Sherry$^7$, Brian Chaboyer, R. A. Koopmann}
\affil{Yale University, Department of Astronomy, P.O. Box 208101, New Haven, CT 06520-8101}
\author{D. Geisler, A. Sarajedini, Arjun Dey}
\affil{National Optical Astronomy Observatories$^1$,
Kitt Peak National Observatory, P.O. Box 26732, Tucson, AZ 85726}
\author{N. Tyson}
\affil{Department of Astrophys. Sciences, Princeton University, Peyton Hall, Princeton NJ 08544}
\author{R. Michael Rich, R. Gal$^8$}
\affil{Department of Astronomy, Columbia University, Mail Code 5242, NY NY 10027}
\author{Robert Lamontagne}
\affil{D\'epartement de Physique and Observatoire du mont M\'egantic, 
Universit\'e de Montr\'eal,
C.P.6128, Succ. Centre-Ville, Montr\'eal, Canada}
\author{N. Caldwell}
\affil{Fred Lawrence Whipple Observatory, P.O. Box 97, 670 Mt. Hopkins Road, Amado, AZ 85645}
\author{Puragra Guhathakurta}
\affil{UCO/Lick Observatory, University of California, Santa Cruz, CA 95064}
\author{A. C. Phillips$^9$, P. Szkody}
\affil{Department of Astronomy, University of Washington, Seattle, WA 98195}
\author{C. Prosser, Luis C. Ho}
\affil{Harvard-Smithsonian Center for Astrophysics 60 Garden Street, Cambridge, MA 02138}
\author{R. McMahan}
\affil{Department of Physics and Astronomy, University of North Carolina,
CB\# 3255 Phillips Hall, Chapel Hill, NC 27599-3255}
\author{G. Baggley}
\affil{Department of Astrophysics, University of Oxford, Keble Road,
                    Oxford, OX1 3RH, England}
\author{K.-P. Cheng}
\affil{California State University, Fullerton, Department of Physics, Fullerton, CA 92634}
\author{R. Havlen}
\affil{Astronomical Society of the Pacific, 390 Ashton Ave., San Francisco, CA 94112}
\author{K. Wakamatsu}
\affil{Physics Department, Gifu University, Gifu, 501-11, Japan}
\author{K. Janes}
\affil{Department of Astronomy, Boston University, 725 Commonwealth Av., Boston, MA 02215}
\author{M. Malkan, F. Baganoff}
\affil{Division of Astronomy, University of California Los Angeles, CA 90095-1562}
\author{P. Seitzer, M. Shara}
\affil{Space Telescope Science Institute, 3700 San Martin Drive, Baltimore, MD 21218}
\author{C. Sturch}
\affil{Computer Sciences Corporation, Space Telescope Science Institute,
Homewood Campus, Baltimore, Maryland 21218}
\author{J. Hesser}
\affil{Dominion Astrophysical Observatory, 5071 West Saanich Road, R.R. 5, Victoria, BC V8X 4M6, Canada}
\author{P. Hartigan$^{10}$, J. Hughes}
\affil{Five College Astronomy Department, University of Massachusetts, Amherst, MA 01003-4525}
\author{D. Welch}
\affil{Department of Physics and Astronomy, McMaster University, Hamilton,
 Ontario L8S 4M1, Canada}
\author{T. B. Williams}
\affil{Department of Physics and Astronomy, Rutgers University, P.O. Box 849,
                    Piscataway, NJ 08855-0849}
\author{H. Ferguson$^{11}$}
\affil{Institute of Astronomy, University of Cambridge, Madingley Road,
                   Cambridge, CB3 OHA, England}
\author{P. J. Francis}
\affil{School of Physics, University of Melbourne, Parkville, Victoria 3052, Australia}
\author{L. French}
\affil{Wheelock College, 200 The Riverway, Boston, MA 02215}
\author{M. Bolte}
\affil{Lick Observatory, University of California, Santa Cruz, CA 95064}
\author{J. Roth}
\affil{Sky Publishing Corporation, 49 Bay State Road, Cambridge, MA, 02138}
\author{S. Odewahn}
\affil{School of Physics and Astronomy, University of Minnesota
          116 Church Street, SE, Minneapolis, MN 55455}
\author{S. Howell}
\affil{Department of Physics and Astronomy, University of Wyoming, Laramie, WY 82071}
\author{W. Krzeminski}
\affil{Carnegie Institution of Washington, Las Campanas Observatory, Casilla 601,
          La Serena, Chile}

\footnoterule

\vspace{0.1in}

\noindent $^1$Cerro Tololo Inter-American Observatory, Kitt Peak National 
Observatory, National Optical Astronomy Observatories, operated by the Association
of Universities for Research in Astronomy, Inc., (AURA), under cooperative agreement
with the National Science Foundation.\\

\noindent $^2$ C\'{a}tedra Presidencial de Ciencias (Chile), 1996-1997.\\

\noindent $^3$Current Address: University of Arizona, Steward Observatory, Tucson, Arizona,
85721\\

\noindent $^4$Current Address: Max-Planck-Institut f\"ur Astronomie, K\"onigstuhl 17,
D-69117 Heidelberg, Germany\\

\noindent $^5$Current Address: Astrophysikalisches Institut Potsdam,
An der Sternwarte 16, D-14482 Potsdam, Germany\\

\noindent $^6$Current Address: European Southern Observatory,
Karl-Schwarzschild-Stra$\beta$e 2, D-85748, Garching b. M\"unchen, Germany\\

\noindent $^7$Current Address: Astronomy Program, S.U.N.Y. at Stony Brook,
Stony Brook, New York.\\

\noindent $^8$Current Address: Caltech, Mail Code 105-24, Pasadena, CA 91125\\

\noindent $^9$Current Address: Lick Observatory, University of California at Santa Cruz,
Santa Cruz, CA 95064\\

\noindent $^{10}$Current Address: Department of Space Physics and Astronomy
Rice University, Houston TX 77251-1892\\

\noindent $^{11}$Current Address: Space Telescope Science Institute, 
3700 San Martin Drive, Baltimore, MD 21218\\

\begin{abstract}
$BV$($RI$)$_{KC}$ light curves are presented for 27 Type~Ia supernovae 
discovered during the course of the Cal\'{a}n/Tololo Survey and for
two other SNe~Ia observed during the same period.  Estimates of the
maximum light magnitudes in the $B$, $V$, and $I$ bands and the initial
decline rate parameter $\Delta$m$_{15}$($B$) are also given.
\end{abstract}


\section{Introduction}

The Cal\'{a}n/Tololo Supernova Survey was begun in 1990 as a collaboration
between astronomers at Cerro Tololo Inter-American Observatory (CTIO) and
the Cerro Cal\'{a}n Observatory of the University of Chile with the principal
goal of examining the Hubble diagram for type Ia supernovae (SNe Ia) out
to redshifts of $\sim$0.1.  During the course of the survey, which we
completed in November 1993, a total of 32 SNe Ia were discovered and
spectroscopically confirmed.  Of these, useful follow-up CCD photometry was
obtained for 27 events.  In addition, as part of the same program, light 
curves were obtained of 2 SNe Ia discovered at other 
observatories. 

In this paper, we present the final reduced $BVRI$ light curves for these 29
SNe Ia, along with estimates of the maximum-light magnitudes in $BVI$ and the
initial decline rate parameter $\Delta$m$_{15}$($B$) (\cite{phillips93}).  Note 
that preliminary light curves for a few events have appeared in 
previous publications (\cite{hamuy93a}, hereafter referred to as Paper I; 
\cite{maza94}, hereafter referred to as Paper II;
\cite{hamuy94}, hereafter referred to as Paper III).
In two accompanying papers, we use these data to
examine 1) the absolute luminosities of the sample (\cite{hamuy96a}, hereafter referred
to as Paper V) and 2) the Hubble diagrams in $BVI$ and value of the Hubble constant
(\cite{hamuy96b} hereafter referred to as Paper VI).

\section{Observations}

The search phase of the Cal\'{a}n/Tololo Supernova Survey consisted of
photographic observations of 45 fields taken with the Curtis Schmidt Camera,
with observations carried out approximately twice a month over the 1990-93 period.
The details of these observations were described in considerable detail
in Paper I, and therefore will not be repeated here.
The follow-up phase consisted
of two parts: 1) classification via optical spectroscopy, and
2) photometric monitoring via direct CCD imaging in the 
$BV$($RI$)$_{KC}$ system.  Of the 50 SNe discovered in the course of the
Survey for which classification spectra were obtained, 32 (64\%) were
found to be type~Ia events.  A complete listing of these SNe~Ia is found
in Table~1 which gives: the SN and host galaxy names, the morphology
and heliocentric redshift of the host galaxy; the line-of-sight extinction
due to our own Galaxy (\cite{bs82}); the 
SN equatorial coordinates derived from an early-epoch CCD image using 
reference stars measured from the digitized
sky survey plates available from the Space Telescope Science Institute;
the offset of the SN from the host galaxy
nucleus; the estimated photographic magnitude of the SN on the discovery plate;
the name of the discoverer; and the UT discovery date.  A V band CCD image
of each SN is reproduced in Figure~1.

Followup photometry
was obtained for as many of these events as proved practical.  Spectra
of three of the SNe (1992O, 1992ai, and 1993af) showed that these
had been caught several weeks or months past maximum light; hence, the 
decision was made not to concentrate on obtaining follow-up photometry of
these events.  For two of the more distant SNe,
1993M and 1993T, an insufficient number of observations were secured 
to provide adequate coverage of the light curves.  Hence, the final
number of Cal\'{a}n/Tololo SNe~Ia for which light curves were 
ultimately obtained was 27.

In addition to these 27 SNe, we obtained $BVRI$ photometry of two other 
SNe~Ia, 1990O and 1992al, which were discovered at other observatories
during the course of the survey.  Information for these two events is 
included at the end of Table~1, and V band CCD images are
shown in Figure~2.  Thus, in this paper we present
light curves for a total of 29 SNe~Ia.

The follow-up photometry for all 29 SNe~Ia was obtained with CCD detectors
on a total of 302 nights between 1990 July 4 and 1995 February 11 thanks
to the extensive collaboration of many visiting astronomers and CTIO
staff members. The vast majority (94\%) of these nights were at CTIO, with the remaining 
observations being carried out with the Las Campanas Observatory (LCO) 
1.0-m telescope and four different telescopes at the European Southern 
Observatory (ESO).  At CTIO, fully 90\% of the data were taken with the
0.9-m telescope, with the remainder coming from the 1.5-m and 4.0-m 
telescopes\footnote[1]{This project serves as an eloquent illustration of the 
capabilities of ``small'' telescopes equipped with state-of-the-art CCD
detectors.}.  A complete journal of the observations is given in Table~2,
which contains the following information: the UT date, the telescope
employed, the observatory, the identity of the CCD detector, and the
observer(s).

\section{Photometric Reductions}

A detailed description of the procedures we followed to produce $BVRI$ 
magnitudes from the individual CCD images of each SN has been given in
Paper III. The various steps are summarized as follows:

\begin{enumerate}

\item Several deep CCD images (in each color) were obtained of the SN field
after the SN had faded from detection.  These images were transformed
geometrically to the same scale, and then coadded to produce a deep
master image of the host galaxy.

\item The master galaxy image was transformed and scaled to the flux
scale of each individual SN image, and then subtracted.  In order to save computing
time, this galaxy subtraction was carried out over
only a subset of the image centered on the SN (see Figure 2 of Paper III).

\item Instrumental magnitudes of the SN and several field local standard
stars were then measured from the galaxy-subtracted images via point spread
function fitting.

\item Finally, the instrumental magnitudes were transformed to the standard
$BV$($RI$)$_{KC}$ system through the use of a photometric sequence set
up in the same field surrounding the SN.  (See Paper I for
further details of the exact photometric transformations employed.)
The photometric sequences for all 29 SNe are identified in the finder
charts in Figures 1 and 2. Only three stars lie outside the observed fields and
could not be identified in these charts, namely: c8 in the field of SN 1990Y which is
located about 130 arcsec west from star c6; c10 in the field of SN 1990Y which is
located about 160 arcsec west and 35 arcsec south from star c6;
and c14 in the field of SN 1990af which is located about 10 arcsec
south and 40 arcsec east from star c12. The magnitudes for the photometric sequences
are listed in Table~3.  In every case, these sequences were derived from
observations made on several (typically 4-6) photometric nights. The uncertainties
quoted correspond to the standard error of the mean.

\end{enumerate}

Table~4 lists the final reduced photometry for each SN.  Please note that 
these magnitudes supersede all previously published values (papers I, II, and III)
for the same SNe. The uncertainties quoted for each magnitude correspond to the sum
in quadrature of the errors due to photon Poisson statistics and an {\it assumed}
additional error of 0.03$^{m}$ in each individual observation. The latter uncertainty
was included in order
to account for errors involved
in the transformation from our instrumental system to the standard system,
and also due to the subtraction of the underlying host galaxy.

\section{Maximum-Light Magnitudes \& Decline Rates}

Figure~3 shows the $BVI$ light curves of the 29 SNe~Ia included in this
study.  Maximum-light magnitudes were derived for each SN in one of the
following two methods:

\begin{enumerate}

\item {\it Direct Measurement.}  For 11 SNe (slightly more than one third of
the sample), photometry was obtained at or before maximum light allowing
direct measurement of the maximum-light magnitudes in $B$ and $V$.
However, for several of these objects (e.g., see the light curves of 
SN~1992ag in Figure 3), coverage of the $I$ light curve was
insufficient to allow direct measurement of the maximum-light brightness
in this band.  In these cases, the best fitting template (see below)
was used, often adjusting this to the first $I$ data point.
The corresponding error in the peak magnitude was taken
to be 0.03$^{m}$ in those cases where the coverage of the light
curve started before maximum, and  0.05$^{m}$ when 
the observations started only one or two days before the peak.

\item {\it Template Fitting.}  For the majority of the SNe in our
sample, the light curve observations did not begin until after
maximum light.  To estimate peak magnitudes for these
events, we employed a template fitting procedure similar to that
utilized in Paper III and \cite{hamuy95} (hereafter referred to as 
Paper IV).  As detailed in a separate paper (\cite{hamuy96c}; hereafter referred
to as Paper VIII), a family of six
$BVI$ light curve templates, representing the range of observed
decline rates of SNe~Ia, were produced from precise CCD
photometry obtained at CTIO of seven well-observed events 
(1992bc, 1991T, 1992al, 1992A, 1992bo, 1993H, and 1991bg).
These templates were fit to the observed photometry of each
of the program SNe via a $\chi^2$-minimizing technique which
solved simultaneously for the time of $B$ maximum and the
peak magnitudes $B_{MAX}$, $V_{MAX}$, and $I_{MAX}$.  (Note
that in our previous papers, the $I$-band data was not
included.)  As detailed in Paper III, before 
performing these fits, the templates were first 
modified by the appropriate K terms (\cite{hamuy93b})
and were also stretched to account for
time dilation.  For about half of the SNe, one of the templates
provided a much better fit (as judged by the value of the
reduced $\chi^2$) than the others.  An example is SN~1992ae
(see Figure~3) whose $BVI$ light curves were found to be
an excellent match to the SN~1992al templates.  However,
for many of the program SNe, the data were fit essentially
equally well by two different templates.  A good example
of such an event is SN~1991ag (see Figure~3), for which 
the 1991T and 1992bc templates yielded similar values 
of the reduced $\chi^2$.  Hence, we adopted the general rule
that when the difference in the reduced $\chi^2$ of two
template fits was $\leq$1.5, the peak magnitudes were obtained
by averaging the results for the two templates.  The
corresponding errors were taken to be the greater of a) half
of the difference between the peak magnitude estimates of the
two templates, b) the 2$\sigma$ formal errors of the $\chi^2$ fits, or
c) 0.05$^{m}$.  When the difference in the reduced $\chi^2$ values
was $>$1.5, the maximum-light magnitudes were taken from the
single-best fitting template, with the adopted error being the
larger of the 2$\sigma$ formal error of the $\chi^2$ fit or 0.05$^{m}$.
Although these rules produced reasonable error estimates in
most cases, we found that the errors derived for
some SNe whose first light curve observations did not begin
until $\sim$2 weeks after maximum were unrealistically low.
Hence, in all cases where template fits indicated that the first 
photometry was not obtained until $\geq$10 days after $B$ maximum,
we adopted the following error estimates: 0.2$^{m}$ in $B$, 0.15$^{m}$
in $V$, and 0.15$^{m}$ in $I$.

\end{enumerate}

For each of the 29 SNe in our sample, we also estimated
the decline rate parameter $\Delta$m$_{15}$($B$) (\cite{phillips93})
which corresponds to the amount in magnitudes that the $B$ light curve 
decreases in brightness during the first 15 days after maximum .
This parameter could
be measured directly for the five best-observed SNe in the sample
(1990af, 1992al, 1992bc, 1992bo, and 1993O).  For the remaining events,
$\Delta$m$_{15}$($B$) was estimated by fitting a parabola to the reduced
$\chi^2$ values yielded by the six template fits (see Paper IV for
further details of this procedure).  Note that when the smallest value 
of the reduced $\chi^2$ corresponded to either of the two extremes of
the range of $\Delta$m$_{15}$($B$) represented by our templates
(SN~1992bc with $\Delta$m$_{15}$($B$) = 0.87 and SN~1991bg with 
$\Delta$m$_{15}$($B$) = 1.93), we set the inferred value of
$\Delta$m$_{15}$($B$) to the same value as the template rather than
attempting to extrapolate a value.

Table~5 summarizes the resulting light curve parameters for all 29 SNe~Ia.
Specifically, we give the epoch of $B$ maximum; the time with respect to
$B_{MAX}$ of the first photometric observation in $B$, $V$, or $I$; the
decline rate parameter $\Delta$m$_{15}$($B$); the apparent maximum-light 
magnitudes in $B$, $V$, and $I$; 
and the method employed to estimate the peak
magnitudes where {\it Data} means that the values were measured directly
from the photometry, {\it Single Fit} indicates that the best fitting
template was used, and {\it Average} signifies that the results from 
the two best template fits were averaged. In Figure~4 we plot a histogram of
the time with respect to $B_{MAX}$ of the first photometric
observation in $B$, $V$, or $I$ for the 29 Cal\'{a}n/Tololo SNe Ia.

In Table~6, we repeat the $\Delta$m$_{15}$($B$) values and give our
final estimates of the peak magnitudes after correction for the 
extinction due to our own Galaxy (see Table~1) and the K~term.
The uncertainties in the corrected magnitudes include errors in the
observed magnitude (listed in Table~5), foreground reddening (0.06$^{m}$ in B, 0.045$^{m}$ in V,
and 0.03$^{m}$ in I), as well as in the K~term (assumed to be $\pm$0.02$^{m}$).
We also list the ``color'' of the SN, $B_{MAX} - V_{MAX}$.
(Note that, strictly speaking, this is not a color since $B_{MAX}$
and $V_{MAX}$ occur at slightly different times.) The uncertainties
in the color were estimated in the following manner: a) for the 11 SNe for
which the photometry started  before maximum light we adopted an error
of 0.03$^{m}$ when the peak was very well observed (6 cases), or 0.05$^{m}$ 
otherwise (5 cases), b) when the coverage of the light curve started between
days 1 and 10 (counted since B$_{MAX}$) the adopted error was the larger 
of half of the difference between the color estimates of the two templates or
0.05$^{m}$; if the single-best fitting template technique was used we adopted an
error of 0.05$^{m}$, or c) when the observations started after day 10 
(counted since B$_{MAX}$) the adopted error was the larger 
of half of the difference between the color estimates of the two templates or
0.10$^{m}$; if the single-best fitting template
technique was used we adopted an 
error of 0.10$^{m}$.

\acknowledgments

This paper was possible thanks to grant 92/0312 from Fondo Nacional de
Ciencias y Tecnolog\'{i}a (FONDECYT-Chile).
MH acknowledges support provided for this work by the National Science Foundation
through grant number GF-1002-96 from the Association of Universities for Research
in Astronomy, Inc., under NSF Cooperative Agreement No. AST-8947990 and from
Fundaci\'{o}n Andes under project C-12984.
JM and MH acknowledge support by C\'{a}tedra Presidencial de Ciencias 1996-1997.

\newpage

\centerline{                             Figure Captions}

\figcaption{V band CCD images of the 32 SNe Ia discovered in the course of the Cal\'{a}n/Tololo survey.
The photometric sequence stars are labeled along with the SNe.}

\figcaption{V band CCD images of the two SNe Ia, 1990O and 1992al, discovered at other observatories
and included in the Cal\'{a}n/Tololo follow-up photometric program.
The photometric sequence stars are labeled along with the SNe.}

\figcaption{B, V, and I light curves for the 29 SNe Ia. In all cases the
solid lines correspond to the best-fitting template. The next-best fit is
also shown as dashed lines when the difference in the reduced $\chi^2$ between the
two best fits was $\leq$1.5.}

\figcaption{Histogram showing the time with respect to $B_{MAX}$ of the first
photometric observation in B, V, or I. Note that for nearly one third of the SNe
the photometric monitoring started before or at maximum light.}


\begin{thebibliography}{}
\bibitem[Burstein \& Heiles 1982]{bs82} Burstein, D., \& Heiles, C. 1982, \aj, 87, 1165
\bibitem[Hamuy et al.\ 1993a]{hamuy93a} Hamuy, M. et al. 1993a, \aj, 106, 2392 (Paper I)
\bibitem[Hamuy et al.\ 1993b]{hamuy93b} Hamuy, M., Phillips, M. M., Wells, L. A.,
  \& Maza, J. 1993b, \pasp, 105, 787
\bibitem[Hamuy et al.\ 1994]{hamuy94} Hamuy, M. et al. 1994, \aj, 108, 2226 (Paper III)
\bibitem[Hamuy et al.\ 1995]{hamuy95} Hamuy, M., Phillips, M. M., Maza, J., Suntzeff, N. B.,
  Schommer, R. A., \& Avil\'{e}s, R. 1995, \aj, 109, 1 (Paper IV)
\bibitem[Hamuy et al.\ 1996a]{hamuy96a} Hamuy, M., Phillips, M. M., Schommer, R.A.,
  Suntzeff, N.B., Maza, J., \& Avil\'{e}s, R. 1996a, \aj, this volume (Paper V)
\bibitem[Hamuy et al.\ 1996b]{hamuy96b} Hamuy, M., Phillips, M. M., Suntzeff, N.B.,
  Schommer, R.A., Maza, J., \& Avil\'{e}s, R. 1996b, \aj, this volume (Paper VI)
\bibitem[Hamuy et al.\ 1996c]{hamuy96c} Hamuy, M., Phillips, M. M., Suntzeff, N.B.,
  Schommer, R.A., Maza, J., Smith, R.C., Lira, P., \& Avil\'{e}s, R. 1996c, this volume (Paper VIII)
\bibitem[Maza et al.\ 1994]{maza94} Maza, J., Hamuy, M., Phillips, M. M., Suntzeff, N. B.,
  \& Avil\'{e}s, R. 1994, \apj, 424, L107 (Paper II)
\bibitem[Phillips 1993]{phillips93} Phillips, M. M. 1993, \apj, 413, L105
\end{thebibliography}
\end{document}